\documentclass[article,aps,showpacs,amssymb]{revtex4}

\begin{document}

\title{The Bianchi identity and weak gravitational lensing}

\author{Thomas P. Kling} \email{tkling@bridgew.edu}

\author{Brian Keith}  \altaffiliation[Sponsored by]{~the Adrian Tinsley
Program for Undergraduate Research, BSC}

\affiliation{Dept. of Physics, Bridgewater State College,
Bridgewater, MA 02325}

\date{\today}

\begin{abstract}

\noindent We consider the Bianchi identity as a field equation for
the distortion of the shapes of images produced by weak
gravitational lensing. Using the spin coefficient formalism of
\textcite{NP}, we show that certain complex components of the Weyl
and Ricci curvature tensors are directly related to fundamental
observables in weak gravitational lensing. In the case of weak
gravitational fields, we then show that the Bianchi identity
provides a field equation for the Ricci tensor assuming a known
Weyl tensor.  From the Bianchi identity, we derive the integral
equation for weak lensing presented by \textcite{miralda95}, thus
making the Bianchi identity a first principles equation of weak
gravitational lensing. This equation is integrated in the
important case of an axially symmetric lens and explicitly
demonstrated in the case of a point lens and a SIS model.

\end{abstract}


\maketitle


\section{Introduction} \label{intro:sec}

\noindent The Bianchi identity in general relativity is usually
expressed as

\begin{equation} \nabla_{[a} {R_{bc]d}}^e = 0 \label{bianchi_gen}
\end{equation}

\noindent and is considered to be a consequence of the geometrical
properties of the covariant derivative operator.  Most students of
relativity come across the Bianchi identity in a contracted form
that is quite useful in discussions of various properties of the
Einstein tensor.

The decomposition of the Riemann tensor into the Ricci ($R_{ab}$)
and Weyl ($C_{abcd}$) tensors,

\begin{equation} R_{abcd} = C_{abcd} + (g_{a[c}R_{d]b} -
g_{b[c}R_{d]a} ) - \frac{1}{3} R g_{a[c} g_{d]b}, \label{decomp}
\end{equation}

\noindent presents a powerful use of the Bianchi identity.  As
\textcite{hawking} point out, the Ricci tensor is determined by
the Einstein field equation, while the Weyl tensor is the part of
the curvature not determined locally by the matter distribution.
However, the application of the Bianchi identity to
Eq.~\ref{decomp} yields a constraint on the Weyl tensor once the
Ricci tensor is found from the Einstein field equations. By
treating the Bianchi identity as a field equation for the Weyl
tensor given a known Ricci tensor, \textcite{NP} (hereafter NP)
developed a consistent methodology for the study of gravitational
radiation.

In this paper, we consider the Bianchi identity in the reverse
manner.  Specifically, we assume that the Weyl tensor is known and
treat the Bianchi identity as a partial differential equation for
the Ricci tensor.

Our motivation for this approach is the study of weak
gravitational lensing, where measurements are made of the
distortion of images due the presence of gravitational fields.
From the observed distortion of images, a matter distribution can
be inferred, so weak gravitational lensing is now a very active
area of observational research with many papers published each
year.  Pivotal early work relating the observed distortion and
inferred mass distribution was done by \textcite{miralda} and
\textcite{KS} amongst others.  Substantial reviews of the subject
are provided by \textcite{mellier} and \textcite{sandb}. For this
paper, we are most interested in the presentation given in
\textcite{miralda95}.

In this paper, we present four results.  First, we show that the
observed distortion and inferred matter distributions of weak
lensing are equal to projected components of the Ricci and Weyl
tensors as expressed in the formalism introduced by \textcite{NP}.
Next, we demonstrate that the integral equation used in
observational weak lensing to determine the projected mass
distribution from observed image distortions is actually an
integral relation between the $\Psi_0$ and $\Phi_{00}$ components
of the Weyl and Ricci tensors.

Our most important result is that we prove that a particular
component of the Bianchi identity yields a partial differential
equation for the $\Phi_{00}$ component of the Ricci tensor given a
known $\Psi_0$ Weyl tensor.  Using a Green's function, we derive
the standard equation used in observational weak lensing as
formulated in \textcite{miralda95} from the Bianchi identity.
Thus, we have developed a PDE approach to weak lensing that may
provide a new calculational approach to the field that allows for
more accurate determinations of mass mappings.

Finally, we consider the important case of axially symmetric
lenses and show that in this case the Bianchi identity is easily
integrated. Two explicit examples are shown: a point lens and a
singular isothermal sphere.

Our results on weak lensing extend a new way of thinking about
gravitational lensing based on a space-time perspective.  Strong
lensing and general image distortion have been studied by several
researchers including  \textcite{fn} and two connected papers by
\textcite{fkn1,fkn2} with a substantial review provided by
\textcite{perlick}.  This paper is the first paper that seeks a
first principles equation for observational weak lensing.


\section{Computation of Ricci and Weyl tensors}
\label{compute:sec}

In this section, we compute the complex Ricci and Weyl tensor
components and the spin coefficients introduced in \citet{NP}. Our
goal is to determine the spin coefficient form of the Ricci and
Weyl tensor components to first order in a static metric
perturbation off flat space.

While modern evidence strongly indicates that the universe is
accelerating and dominated by dark energy, it is reasonable to
chose to perturb off flat space for several reasons. First, we
wish to directly compare our derivation of the equations of weak
lensing with the standard derivations using the thin-lens
approximation. Because all the ``lensing activity'' in the
thin-lens approximation happens when light rays passes the lens,
the choice of cosmology does not directly enter the equations.  In
fact, both in the literature and in actual application, the only
place the cosmological model enters weak lensing calculations is
in the conversion between redshift and angular diameter distance.
Hence, one of the strengths of lensing studies is that the basic
equations of the theory are relatively independent of cosmological
model.

Furthermore, since the basic cosmological models are conformally
flat, our basic results can be translated to them. Beginning with
a more ``realistic'' metric does not add to the calculation and
introduces a number of conceptual issues best handled in a
different paper.

The physical situation that we wish to study is the appearance and
distortion of a small patch of the sky containing extended
galaxies that has been weakly lensed by some matter distribution.
 Both the astronomer in question and the sky of galaxies are
assumed to be situated in reasonably isolated regions of
space-time where the space-time is flat.

The observed galaxies are connected to our astronomer by light
rays along her past light cone.  In fact, we may think of pencils
of light rays that travel from each extended galaxy to the
telescope. The path of these light rays is to be determined by
integrating the null geodesic equations of the space-time metric,
so that we may find a tangent vector, $\ell^a$, to the light rays.

The matter distribution that lenses the light rays is encoded into
a space-time metric.  For our purposes, we may assume a weakly
perturbed Minkowski metric,

\begin{equation} ds^2 = (1+2\varphi)dt^2 - (1-2\varphi)(dx^2 + dy^2
+dz^2), \label{metric} \end{equation}

\noindent where $\varphi$ is a static gravitational potential
satisfying

\begin{equation} \nabla^2 \varphi = 4 \pi \rho(x,y,z).
\label{laplace} \end{equation}

\noindent As the astronomer and observed galaxies are to be
isolated, we assume that the matter distribution, $\rho(x,y,z)$,
is contained in some region of space far from the astronomer and
galaxies.  In practice, this is always the case.

Following the program of NP, we chose a tetrad of null vectors,

\begin{equation} \lambda_i^a = (\lambda_1^a, \lambda_2^a,
\lambda_3^a, \lambda_4^a) = ( \ell^a, n^a, m^a, \bar m^a),
\label{tetrad_def} \end{equation}

\noindent associated with the pencil of light rays connecting our
astronomer and each individual observed galaxy, where $\ell^a$ is
the real vector tangent to the pencil.  By convention, the other
vectors are chosen such that

\begin{equation} \ell^a n_a = 1 \quad\quad \& \quad\quad m^a \bar m_a =
-1, \label{tetrad_conds} \end{equation}

\noindent with all other products zero. $m^a$ and $\bar m^a$ are
spatial, complex null vectors parallel propagated along the
pencil.

In principle, the space-time metric can be written in terms of a
general null tetrad as $g^{ab} = \eta^{ij} \lambda^a_i
\lambda^b_j$, although this is not central to our presentation.
Further, we will make physical restrictions on the tetrad vectors
tied to our goal of finding the Weyl and Ricci tensor components
to first order in the perturbation.

We orient our spatial coordinates, $(x,y,z)$, such that the
astronomer is located at $x = y = 0$ and $z = z_a$ with the
``center'' of the matter distribution at the spatial origin. With
this placement, our astronomer's telescope points straight down
the $\hat z$ axis.

In principle, each null tetrad associated with an individual
pencil of light connecting the astronomer to a lensed galaxy
varies along the pencil.  However, in our calculations, we are
justified in considering each tetrad to be a constant null tetrad
given by

\begin{equation} \ell^a = \frac{1}{\sqrt{2}} \left(1, 0, 0, 1
\right), \quad\quad  n^a = \frac{1}{\sqrt{2}} \left( 1, 0,0, -1
\right) \quad \& \quad  m^a =  \frac{1}{\sqrt{2}} \left( 0, 1 , i,
0 \right), \label{tetrad} \end{equation}

\noindent for two reasons.  First, the wide field telescopes used
today see a very small portion of the sky.  For example, the
proposed $8.4$~m Dark Matter Telescope would have a field of view
of only $260$ square milli degrees \cite{DMT}.  Hence, with our
orientation, all the pencils of light are parallel to the $\hat z$
axis.  Second, even though the individual pencils will be
deflected by the lens, the deflection of the light ray will be
proportional to the 2-dimensional $(x,y)$ gradient of $\varphi$.
 Since we will be contracting the tetrad vectors with the Ricci and
Weyl tensors, to work to first order in $\varphi$, we must neglect
any variation in the null tetrad.

The basis of the NP formalism is twelve complex spin coefficients
that are analogous to the twenty-four real Ricci rotation
coefficients defined by

\begin{equation} \gamma^i_{jk} = \lambda_j^a \lambda_k^b \nabla_b
\lambda^i_a. \label{spin} \end{equation}

\noindent Because our choice of tetrad involves vectors whose
components are all constant, all the NP spin coefficients are zero
for our physical situation (to zeroth order in the metric
perturbation).

In the NP formalism, tetrad components of the Ricci and Weyl
tensors are computed by contracting the coordinate components with
the tetrad vectors.  This yields five complex Weyl tensor and ten
complex Ricci tensor components (see appendix \ref{app:sec}).

Using the null tetrad above, the non-zero NP formalism Ricci
curvature components are

\begin{equation} \Phi_{00} = \Phi_{22} = 2\Phi_{11} = \frac{1}{2}
\nabla^2 \varphi  . \label{ricci_comps} \end{equation}

\noindent The first order Weyl tensor components are

\begin{eqnarray} \Psi_0 = \Psi_4 &=& \frac{1}{2} \left( \varphi_{xx}
- \varphi_{yy} + 2i\varphi_{xy} \right) , \nonumber \\ \Psi_1 =
\overline{\Psi}_3 & = & - \frac{1}{2} \left( \varphi_{xz} + i
\varphi_{yz} \right) , \nonumber \\ \Psi_2 & =& \frac{1}{2} \left(
\varphi_{zz} - \frac{1}{3} \nabla^2 \varphi \right) .
\label{weyl_comps} \end{eqnarray}

\noindent We note that up to numerical factors, $\Phi_{00}$ is
equal to the matter density $\rho (x,y,z)$ that determines the
gravitational perturbation $\varphi$.

The four directional derivatives associated with the null tetrad
play an important role in the NP formalism and are given special
names:

\begin{equation} D = \ell^a \partial_a,  \quad\quad \Delta = n^a
\partial_a , \quad\quad \delta = m^a \partial_a,  \quad \&
\quad \bar\delta = \bar m^a \partial_a .\label{derivs1}
\end{equation}

\noindent Using our choice of tetrad, and the fact that the metric
perturbation is static, these derivative operators will act as

\begin{eqnarray} & D = \frac{1}{\sqrt{2}} \, \frac{
\partial}{\partial z} ,  \quad\quad\quad  \Delta = - \frac{1}{\sqrt{2}} \, \frac{
\partial}{\partial z} , & \nonumber \\  & \delta = \frac{1}{\sqrt{2}} \,
\left( \frac{\partial}{\partial x} + i \frac{\partial}{\partial y}
\right) ,  \quad\quad\quad  \bar \delta = \frac{1}{\sqrt{2}} \,
\left( \frac{\partial}{\partial x} - i \frac{\partial}{\partial y}
\right). & \label{derivs2} \end{eqnarray}


\section{Observational weak lensing in practice} \label{obs_lensing:sec}

In this section, we briefly outline the usual presentation of the
equations of weak lensing by a thin lens so that we can draw
parallels to our presentation that uses the Weyl and Ricci tensor.
The derivation presented here relies heavily on that of
\textcite{miralda95}.

The practicing astrophysicist assumes a thin (or two dimensional)
lens that divides a background space-time into an observer and a
source side. The lens lies in a two (spatial) dimensional ``lens
plane'' that is perpendicular to the line of sight of the
telescope.  The observed background galaxy is said to lie in a two
dimensional ``source plane,'' also perpendicular to the line of
sight.

Using dimensionless cartesian coordinates $\vec{r}_s$ in the
source plane and $\vec{r}_l = (x,y)$ for the lens plane, the
starting point for thin-lens lensing is a mapping from the lens
plane to the source plane given by

\begin{equation} \vec{r}_s = \vec{r}_l - \vec{\nabla}_2 \,
\psi(\vec{r}_l). \label{map} \end{equation}

\noindent The two-dimensional gravitational potential $\psi$ is
determined from an ordinary three-dimensional potential $\varphi$
by projection,

\begin{equation} \psi(\vec{r}_l) = \int_{z_a}^{z_s} \, dz
\, \varphi (x,y,z), \label{2dgrav} \end{equation}

\noindent where $z_a$ and $z_s$ are the $z$ coordinates of the
observing astronomer and distant source, respectively.  The two
dimensional potential $\psi$ will satisfy

\begin{equation} \frac{1}{2} \Delta^2 \psi = \frac{1}{2} \left(
\partial_x^2 + \partial_y^2 \right) \psi =
\frac{\Sigma(\vec{r}_l)} {\Sigma_{crit}}, \label{lap2d}
\end{equation}

\noindent where $\Sigma(\vec{r}_l)$ is a projected mass density
and $\Sigma_{crit}$ is a critical density for strong lensing, or
the appearance of multiple images.

To consider weak lensing, one forms a Jacobian matrix

\begin{equation} {\cal{J}} = \frac{\partial \vec{r}_s} {\vec{r}_l}
= \left( \begin{array}{cc} 1 - \psi_{xx} & -\psi_{xy} \\
-\psi_{xy} & 1 - \psi_{yy} \end{array} \right) = \left(
\begin{array}{cc} 1 - \kappa - \lambda & -\mu \\
-\mu & 1 - \kappa + \lambda \end{array} \right), \label{j_tl}
\end{equation}

\noindent where

\begin{equation} \kappa = \frac{1}{2} \left( \psi_{xx} + \psi_{yy}
\right) = \frac{1}{2} \Delta^2 \psi = \frac{\Sigma}{\Sigma_{crit}}
\label{k1} \end{equation}

\noindent is referred to as the ``convergence,'' and the two
quantities

\begin{equation} \lambda = \frac{1}{2} \left( \psi_{xx} - \psi_{yy}
\right) \quad\quad \& \quad\quad \mu = \psi_{xy} \label{shears}
\end{equation}

\noindent are called the ``shears'' in the thin-lens literature.
\textcite{fkn2} shows how these quantities are related to the
convergence ($\rho$) and shear ($\sigma$) in general relativity.
For this paper, it is critical to note the similarities that
$\kappa$ has with $\Phi_{00}$ and that $\Psi_0$ has with $\lambda$
and  $\mu$.

If the outer surface of a circular, extended source is
parameterized by $\vec{t}_s = R \left( \cos \delta, \sin \delta
\right)$, then under the thin lens mapping the inverse of the
Jacobian matrix in Eq.~\ref{j_tl} maps $\vec{t}_s \rightarrow
\vec{t}_l$, a new elliptical curve in the lens plane. The
orientation and ellipticity of the resulting ellipse is determined
by $\lambda$ and $\mu$ -- see \textcite{ehlers} for the details.

Since the orientation and ellipticity is observable, the goal in
observational weak lensing is to measure the shears $(\lambda, \,
\mu)$ and infer the projected mass density, $\kappa$.  To do this,
one inverts Eq.~\ref{k1} using the two dimensional Green's
function and applies the differentiations in Eq.~\ref{shears} to
the result.  This yields

\begin{eqnarray} \lambda & =& \int \, d\vec{r}' \, \frac{\kappa
(\vec{r}')}{\pi} \, \frac{ (-\cos 2\eta )}{ | \vec{r} - \vec{r}'
|^2 }, \nonumber \\ \mu & =& \int \, d\vec{r}' \, \frac{\kappa
(\vec{r}')}{\pi} \, \frac{ (-\sin 2\eta )}{ | \vec{r} - \vec{r}'
|^2 }, \label{k3} \end{eqnarray}

\noindent where $\eta$ is the angle in the lens plane between
$\vec{r}-\vec{r}'$ and the $\hat x$ axis.

By employing a Fourier transform technique, \textcite{miralda95}
shows that one can invert Eq.~\ref{k3} to obtain

\begin{equation} \kappa (\vec{r}) = \int \, d\vec{r}'
\, \frac{[\lambda(\vec{r}') \, , \, \mu(\vec{r}') ]}{\pi} \cdot
\frac{ [ \cos 2\eta \, , \, \sin 2\eta ]}{ | \vec{r} - \vec{r}'
|^2} . \label{k4} \end{equation}

\noindent Equation~\ref{k4} is the primary equation for
observational weak lensing.  It is an integral relation that
allows one to determine the projected mass density at every point
if one has measured the ``shears.''


\section{Weak lensing with curvature tensors} \label{obs_gr:sec}

We may exploit the similarities between the definitions of the
thin-lens convergence and shears and $\Phi_{00}$ and $\Psi_0$ to
recast Eq.~\ref{k4} in different language.  The first step is to
project our gravitational potential and curvature tensors into a
``thin-lens'' form.  Equation~\ref{2dgrav} projects our metric
perturbation $\varphi(x,y,z)$ to a two dimensional $\psi(x,y)$.
Then we define

\begin{equation} _L\Phi_{00} \equiv  \int_{z_a}^{z_s} \, dz
\, \Phi_{00} = \frac{1}{2} \left( \psi_{xx} + \psi_{yy} \right),
\label{lens_phi00} \end{equation}

\noindent where we use Eq.~\ref{ricci_comps} with Eq.~\ref{2dgrav}
and the property that $\varphi_z$ is zero far from the lens.
Likewise, we can define

\begin{equation} _L\Psi_0 \equiv  \int_{z_a}^{z_s} \, dz
\, \Psi_0 = \frac{1}{2} \left( \psi_{xx} - \psi_{yy} + 2i
\psi_{xy} \right). \label{lens_psi0} \end{equation}

We note that $_L\Phi_{00} = \kappa$ and $_L \Psi_0 = \lambda + i
\mu$, so that the fundamental observable in weak gravitational
lensing is a projected component of the Weyl tensor, while the
fundamental inferred quantity of observational interest (the
projected mass density) is a projected component of the Ricci
tensor.

Then exactly as in section~\ref{obs_lensing:sec}, one can use a
two-dimensional Green's function to invert Eq.~\ref{lens_phi00}
for $\psi$ and plug this relation into Eq.~\ref{lens_psi0} to
obtain an integral equation specifying $_L \Psi_0$ given a known
$_L \Phi_{00}$.  Taking the Fourier transform yields a relation
between the transforms of $_L \Psi_0$ and $_L \Phi_{00}$ which is
easily rearranged such that the inverse Fourier transform produces
the desired integral equation for $_L \Phi_{00}$,

\begin{equation} _L \Phi_{00} \, (\vec r) = - \int \, d\vec{r}' \, \frac{_L
\Psi_0 (\vec{r}')}{\pi} \, \frac{e^{-2i\eta}}{|\vec{r} -
\vec{r}'|^2}, \label{int_relation} \end{equation}

\noindent where $\eta$ again is the angle between $\vec{r} -
\vec{r}'$ and the $\hat x$ axis in the lens plane.

Equation~\ref{int_relation} demonstrates that the fundamental
equation of weak lensing is an integral relation between
components of the Weyl and Ricci curvature tensors.  Specifically,
a projected version of $\Psi_0$ is known through observation and
used to determine a projected mass density $_L \Phi_{00}$.


\section{The Bianchi identity} \label{bianchi:sec}

As discussed in section~\ref{intro:sec}, applying the Bianchi
identity, Eq.~\ref{bianchi_gen}, to the decomposition of the
Riemann tensor into the Ricci and Weyl curvature tensors produces
a differential constraint on the Weyl tensor.  To obtain the
Bianchi identity in the NP spin coefficient formalism, one
contracts the Bianchi identity with all possible combinations of
the tetrad vectors and writes the resulting equations out using
the definitions of the spin coefficients and Ricci and Weyl tensor
components. In the non-vacuum case, this results in twelve
equations, which are listed in a number of references
including~\textcite{NT}.

The first full Bianchi identity equation is

\begin{eqnarray} \bar\delta \Psi_0 - D \Psi_1 + D \Phi_{01} -
\delta \Phi_{00} &= & ~ (4\alpha - \pi) \Psi_0 -
2(2\rho+\epsilon)\Psi_1 + 3\kappa \Psi_2 + (\bar \pi - 2\bar
\alpha - 2\beta) \Phi_{00} \nonumber \\ & & ~~~ + 2(\epsilon +
\bar\rho) \Phi_{01} + 2\sigma \Phi_{10} -2\kappa \Phi_{11} -
\bar\kappa \Phi_{02}. \label{bianchi1} \end{eqnarray}

\noindent Due to our choice of tetrad, all the spin coefficients
are zero, so the entire right hand side of Eq.~\ref{bianchi1}
vanishes. Also, $\Phi_{01}$ is zero, so to first order in the
gravitational perturbation, we have

\begin{equation} \bar\delta \Psi_0 - D \Psi_1 - \delta \Phi_{00} =
0. \label{bianchi2} \end{equation}

Equation~\ref{bianchi2} is a partial differential equation that
holds at every point along the pencil of rays connecting our
astronomer with the distant observed galaxy.  To compare this
relation with Eq.~\ref{int_relation}, we project into a lens plane
by integrating out the $z$ coordinate.  Integrating $D \Psi_1$ in
$z$ from $z_a$ to $z_s$ effectively yields zero since $D \sim
\partial_z$ and $\Psi_1$ evaluated far from the lens is assumed to be zero.
Passing the $z$ integration through the $\delta$ derivative
operators, we have

\begin{equation} \delta \, _L \Phi_{00}  = \bar\delta \, _L\Psi_0.
\label{bianchi3} \end{equation}

We consider Eq.~\ref{bianchi3} as a field equation for $_L
\Phi_{00}$ given a known $_L \Psi_0$. Equation~\ref{bianchi3}
relates the fundamental quantities of weak gravitational lensing
and is derived from first principles in general relativity.


\section{Green's function}

In this section, we show that the relation between the Weyl and
Ricci tensor derived from the Bianchi identity,
Eq.~\ref{bianchi3}, is the differential version of the integral
relation used in standard weak lensing studies,
Eq.~\ref{int_relation}.  To do this, we employ a set of Green's
functions first developed by \citet{porter} and expanded in
\citet{ivancovich}, which we refer to as the Porter Green's
functions.

The Porter Green's functions are designed to work with powers of
the $\eth$ and $\bar\eth$ differential operators.  The $\eth$
derivative operator acts on functions on the sphere of different
spin weight in different ways and serves as a spin-weight raising
and lowering operator \cite{NT}.  Functions on the sphere of a
spin weight $s$ are expandable in a series of spin-weighted
spherical harmonics $_sY_{lm}$.  Because different powers of
$\eth$ and $\bar\eth$ annihilate the $_sY_{lm}$ with $|s|=l$, a
set of Green's functions can be developed.

Our main result, Eq.~\ref{bianchi3}, is a partial differential
equation that holds in an $(x,y)$ plane that we consider to be the
lens plane.  To make contact with the Porter Green's functions, we
introduce the pair of complex coordinates $(\zeta, \bar\zeta)$
defined by

\begin{equation} \zeta = \frac{1}{\sqrt{2}}(x - i y) . \label{stereo}
\end{equation}

\noindent In these coordinates, $m^a$ points in the $\zeta$
direction and $\delta = \partial_\zeta$.

By stereographic projection of the sphere into a equatorial plane
from the pole, the $(\zeta, \bar\zeta)$ coordinates are complex
coordinates for the sphere. With the exception of the point at
infinity, where the projected curvature tensors $_L\Psi_0$ and
$_L\Phi_{00}$ are assumed to be zero, introduction of the complex
stereographic coordinates turns Eq.~\ref{bianchi3} into a partial
differential equation on the sphere.

Since $_L\Phi_{00}$ is a spin-weight zero function, the
application of $\eth$ to it will take the form

\begin{equation} \eth \, _L\Phi_{00} = (1 + \zeta \bar \zeta ) \delta
_L\Phi_{00}. \label{e1} \end{equation}

\noindent For this reason, we multiply both sides of
Eq.~\ref{bianchi3} by $(1+\zeta\bar\zeta)$ and our Bianchi
identity takes the form

\begin{equation} \eth \, _L \Phi_{00} = (1+\zeta \bar \zeta)
\bar \delta \, _L \Psi_0 = A_1(\zeta,\bar\zeta), \label{b2}
\end{equation}

\noindent where $A_1 (\zeta, \bar\zeta)$ is a function of
spin-weight 1.

The Porter Green's function for an equation of the form in
Eq.~\ref{b2} is

\begin{equation} _L\Phi_{00} (\zeta,\bar\zeta) = \int_{S^2} \,
K_{0,-1}(\zeta,\bar\zeta;\eta,\bar\eta) \, (1+\eta\bar\eta) \,
\bar\delta_\eta \, _L\Psi_0 d\mu_\eta, \label{b3} \end{equation}

\noindent where the integral is taken over the sphere
parameterized by complex coordinates $(\eta, \bar\eta)$ with the
area element

\[ d\mu_\eta = \frac{2}{i} \frac{d\eta \wedge d\bar\eta}
{(1 + \eta \bar \eta)^2}. \]

\noindent The kernel of the Green's function for an equation of
the type in Eq.~\ref{b2} is

\begin{equation} K_{0,-1}(\zeta,\bar\zeta;\eta,\bar\eta)  =
\frac{1}{4\pi} \frac{1 + \eta\bar\eta}{\bar \zeta - \bar \eta} .
\label{kernel} \end{equation}

Putting all this together, we have

\begin{equation} _L\Phi_{00} (\zeta)  = \frac{1}{2 i \pi} \int_{S^2}
\frac{\bar\delta_\eta \, _L \Psi_0 (\eta)}{\bar\zeta - \bar \eta}
d \eta \wedge d \bar\eta. \label{b4} \end{equation}

\noindent It is convenient to multiply by $1$ in the form $1 =
(\zeta - \eta)/(\zeta-\eta)$ to obtain

\begin{equation} _L\Phi_{00} (\zeta)  = \frac{1}{2 i \pi} \int_{S^2}
\frac{\zeta - \eta}{(\bar\zeta - \bar \eta) (\zeta-\eta) }
\bar\delta_\eta \, _L \Psi_0 (\eta) d \eta \wedge d \bar\eta.
\label{b5} \end{equation}

Because we want to show that Eq.~\ref{b5} is equivalent to
Eq.~\ref{int_relation}, we reintroduce cartesian coordinates

\begin{equation} \zeta = \frac{1}{\sqrt{2}} (x-i y) \quad\quad \eta =
\frac{1}{\sqrt{2}} (x' - i y'). \label{coords} \end{equation}

\noindent In these coordinates,  Eq.~\ref{b5} becomes

\begin{equation} _L\Phi_{00} (x,y) = \frac{1}{2\pi} \int
\frac{[(x-x')-i(y-y')]}{(x-x')^2+(y-y')^2} \left( \partial_{x'} -
i \partial_{y'} \right) ~_L \Psi_0 \, dx'dy' .\label{b6}
\end{equation}

Comparing Eq.~\ref{b6} with Eq.~\ref{int_relation}, we see that we
need to flip the $\bar\delta_\eta$ derivative operator that is
inside the integral by integrating by parts.  Using the definition
of the angle $\eta$  as the angle the vector $\vec{r} - \vec{r}'$
makes with the $+\hat x$ axis and integrating Eq.~\ref{b6} by
parts yields

\begin{equation} _L\Phi_{00} (x,y) = - \int \, d\vec{r}' \, \frac{_L\Psi_0}{\pi}
\, \frac{e^{-2i\eta}}{|\vec{r} - \vec{r}'|^2} \label{bfinal}
\end{equation}

\noindent plus the surface term

\begin{equation}  \frac{1}{2 \pi} \int \, dx'
dy' \left( \partial_{x'}  -i \partial_{y'} \right) \left\{
\frac{[(x-x') - i(y-y')]}{|\vec{r} - \vec{r}'|^2} ~_L\Psi_0
\right\}. \label{surface} \end{equation}

\noindent One can see that the surface term will vanish by
breaking the integral into two parts for the $\partial_{x'}$ and
$\partial_{y'}$ derivatives.  Each part vanishes as the integrand
is evaluated at either $x'=\infty$ or $y'=\infty$.

Therefore, we have shown that the application of the Porter
Green's function to  the Bianchi identity, Eq.~\ref{bianchi3},
results in the integral relation used in weak gravitational
lensing studies.


\section{Axial Symmetry} \label{axial:sec}

In this section, we consider those mass distributions which are
axially symmetric around the line of sight.  In practice, this is
a very important simplifying assumption used in gravitational
lensing studies.  Under this assumption, Eq.~\ref{bianchi3} is
easily integrated directly, without using the Porter Green's
function.

First, we change coordinates in the lens plane from the cartesian
$(x\, , \, y)$ coordinates to standard axial coordinates $(r \, ,
\, \phi)$.  In the axial coordinate system, the $\delta$
directional derivative operator in Eq.~\ref{derivs2} is

\begin{equation} \delta = \frac{e^{i \, \phi}}{\sqrt{2}} \left(
\partial_r + \frac{i}{r} \partial_\phi \right). \label{derivs3}
\end{equation}

As an ansatz for an axially symmetric lens, we assume that the
Ricci tensor is a function of $r$ only,

\begin{equation} ~_L \Phi_{00} = ~_L \Phi_{00} (r),
\label{ansatz1} \end{equation}

\noindent and that the Weyl tensor can be written as

\begin{equation} ~_L \Psi_0 = \Upsilon(r) \, e^{2i \, \phi}.
\label{ansatz2} \end{equation}

\noindent This second assumption is motivated by the concept of
spin weight in the NP formalism \cite{NT}.

Applying Eq.~\ref{derivs3} to our ansatz for the solution to
Eq.~\ref{bianchi3}, the $\phi$ dependence drops out, and one is
left with

\begin{equation} \partial_r \, _L \Phi_{00} (r) = \partial_r
\Upsilon (r) \, +\,  \frac{2}{r} \, \Upsilon (r), \label{bianchi4}
\end{equation}

\noindent which is now a one dimensional equation.

Recall that we have shown that $_L \Psi_0$ or $\Upsilon (r)$ is a
known, measurable quantity and that $_L \Phi_{00}$ is essentially
the projected matter density. This means that after integrating
over $r$, Eq.~\ref{bianchi4} gives us the functional form of the
projected matter distribution:

\begin{equation} _L \Phi_{00} (r) ~=~ \Upsilon (r) ~+~ \int \, \frac{2}{r}
\, \Upsilon \, dr + C. \label{bianchi5} \end{equation}

\noindent The constant of integration represents the mass sheet
degeneracy present in gravitational lensing and can be taken as
zero.


\section{Two axially symmetric examples} \label{models:sec}

In this section, we examine two axially symmetric mass
distributions in the context of weak lensing as expressed by
Eq.~\ref{bianchi5}. The models we consider include a singular,
Schwarzschild-like point lens and a singular isothermal sphere
(SIS) model. For each model, we will determine the gravitational
potential, $\psi (r)$, then compute the projected Weyl and Ricci
tensor components and show that they obey Eq.~\ref{bianchi5}.

\subsection{Point lens}

As our first example, we consider a Schwarzschild point lens of
mass $M$ at the origin.  For this mass distribution, the
gravitational potential $\psi$ is given by

\begin{equation} \psi = 2 M \, \ln (r) = 2 M \, \ln \left(
\sqrt{x^2 + y^2} \right). \label{m1grav} \end{equation}

\noindent  One can show by direct computation that

\begin{equation} ~_L\Psi_0 = -\frac{2\,M}{r^2}\, e^{2i\,\phi},
\label{m1Psi} \end{equation}

\noindent which matches Eq.~\ref{ansatz2} yielding

\begin{equation} \Upsilon = -\frac{2 \, M}{r^2}. \label{m1Up}
\end{equation}

Inserting this functional form for $\Upsilon$ into the right hand
side of Eq.~\ref{bianchi5} gives

\begin{equation} \frac{-2 \, M}{r^2} + \int~ dr \frac{-4 \,
M}{r^3}~=~ 0. \label{m1rhs} \end{equation}

\noindent Equation~\ref{m1rhs} indicates, through the Bianchi
identity Eq.~\ref{bianchi5}, that the projected Ricci tensor is
zero at all radii $r>0$, which is the expected result for a
Schwarzschild, point-like lens.

\subsection{SIS model}

The SIS model is a one parameter model given by

\begin{equation} \Sigma (r) = \frac{\sigma_v^2}{2 \, r} ,
\label{SIS} \end{equation}

\noindent where $\sigma_v$ is the velocity dispersion. Although
unphysical due to the singularity at the origin and infinite total
mass, SIS models do explain the flat rotation curves of galaxies
and are widely used as models in gravitational lensing studies.

For our purposes, the simplest way to integrate the projected
gravitational potential for the SIS model is to follow the
argument of \citet{ehlers} in section 8.1, where it is shown that
one can write the projected gravitational potential as

\begin{equation} \psi (r) = 8 \pi \, \int_0^r \, r' \, dr' \,
\Sigma\,(r') \ln\left( \frac{r}{r'} \right). \label{2dgreen2}
\end{equation}

\noindent It is then relatively simple to show that

\begin{equation} \psi = 4 \pi \, \sigma_v^2 \, r , \label{m2grav}
\end{equation}

\noindent so that the projected Weyl tensor is

\begin{equation} ~_L\Psi_0 = - \frac {2 \pi \, \sigma^2_v}{r} \,
e^{2i \, \phi}. \label{m2Psi} \end{equation}

From Eq.~\ref{ansatz2}, we have $\Upsilon = - 2 \pi \sigma_v^2 /
r$ and from the Bianchi identity

\begin{equation} \Upsilon + \int \, dr \, \frac{2\Upsilon}{r} =
\frac{2 \pi \, \sigma_v^2}{r}. \label{m2rhs} \end{equation}

\noindent By direct computation,

\begin{equation}  ~_L\Phi_{00} = \frac{1}{2} \left( \psi_{xx} +
\psi_{yy} \right) = \frac{2 \pi \, \sigma_v^2}{r}, \label{m2lhs}
\end{equation}

\noindent so that the Bianchi identity, Eq.~\ref{bianchi5} is
preserved.


\section{Discussion}

The main results of this paper are that the fundamental quantities
of weak gravitational lensing are projected components of the
Ricci and Weyl tensors and that the Bianchi identity provides a
field equation for the projected matter density derived from first
principles.  We explicitly show that the common integral equation
used in weak gravitational is an integral version of the Bianchi
identity.  To our knowledge, these results are known in the
lensing or relativity communities.

From this perspective, this paper extends the work presented in a
series of recent papers that has helped reunite applied
gravitational lensing with its roots in general relativity. The
underlying perspective of these papers has been that any observed
lensing phenomena must be encoded into the past light cone of the
observer.

A primary purpose of these papers has been to identify the
equation from general relativity that is approximated in the thin
lens treatment used by practicing astrophysicists.  \textcite{fn}
first pointed out that the lens should be coded into a space-time
metric, and then solving the null geodesic equations would
simultaneously give the ``time of flight'' equation and lens
mapping.  In a two-paper set, \textcite{fkn1} related general
relativity's optical scalars to the ``shears'' and
``convergences'' cited in the thin-lens literature.  These papers
also wrote out integral relations that showed how image distortion
grew continuously along the pencil of rays connected a source and
observer. A generalization of the Fermat principle of least time
was present in \textcite{fkn3}.

Because Eq.~\ref{int_relation} is kinematically derived, it does
not indicate a true connection between observational weak lensing
and relativistic first principles.  The derivation of
Eq.~\ref{int_relation} relies only on the definitions of
$\Phi_{00}$ and $\Psi_0$.

However, we show in this paper that the projected Bianchi
identity, Eq.~\ref{bianchi3} is the differential version of
Eq.~\ref{int_relation}.  Thus, we have a first principles
derivation of the basic equation of weak gravitational lensing for
the first time and a potentially useful new PDE approach to the
subject.

Most directly, Eq.~\ref{bianchi3} provides a new PDE approach for
determining the projected mass density in weak lensing studies.
While numerical approaches to solving PDEs with real data tend to
less stable than integral approaches, they also tend to provide
higher accuracy, so it is not unreasonable to pursue approaches to
data analysis based directly on these results.

Another potential application of this work might be in the
cumulative analysis of thick lenses or repeated lensing by
multiple clusters along the line of sight. In this case, one would
not apply a thin-lens assumption, but would use Eq.~\ref{bianchi2}
as a field equation along the line of sight for a full metric that
encodes a more realistic cosmological model.


\begin{acknowledgments}
The authors wish to acknowledge Simonetta Frittelli, Ted Newman
and Al Janis for helpful suggestions and comments.  BK would like
to thank the Adrian Tinsley Program for Undergraduate Research of
BSC for making his involvement possible.
\end{acknowledgments}

\appendix

\section{Weak field Ricci and Weyl tensors} \label{app:sec}

In this appendix, we give the usual component form of the Ricci
and Weyl tensors considered in the text, along with their
definitions in the \citet{NP} formalism.  In
Eqs.~\ref{christofel}-\ref{weyl}, the usual Einstein summation
conventions are not employed.

For the weak field metric in Eq.~\ref{metric}, to first order in
the perturbation, $\varphi$, the non-zero Christofel symbols are

\begin{eqnarray} {\Gamma^0}_{0i} = \varphi_i & \quad\quad &
{\Gamma^i}_{ik} = -\varphi_k \nonumber \\ {\Gamma^i}_{00} =
\varphi_i & \quad\quad & {\Gamma^i}_{kk} = \varphi_i \quad (i \ne
k) \label{christofel} \end{eqnarray}

\noindent where $\varphi_i = \partial_i \varphi$, $0$ denotes time
and $i,j,k$ denote spatial components. The non-zero, first order
Ricci tensor components are

\begin{equation} R_{00} = - \nabla^2 \varphi \quad\quad R_{ii} =
-\nabla^2 \varphi, \label{ricci} \end{equation}

\noindent with $\nabla^2 = \partial_x^2 + \partial_y^2 +
\partial_z^2$, while the non-zero, first order Weyl tensor components
are

\begin{eqnarray} C_{0i0i} = \frac{1}{3} \left( -3 \varphi_{ii} +
\nabla^2 \varphi \right) & \quad\quad & C_{0i0j} = -\varphi_{ij}
\quad i \ne j \nonumber \\ C_{ijij} = \frac{1}{3} \left( 3
\varphi_{kk} - \nabla^2 \varphi \right) \quad i\ne j \ne k &
\quad\quad & C_{ijik} = -\varphi_{jk} \quad i \ne j \ne k.
\label{weyl} \end{eqnarray}

In the NP formalism, the components of the Ricci and Weyl tensors
are contracted with the null tetrad to give named components. The
Weyl tensor components are defined by

\[\Psi_0 = -C_{abcd} \ell^a m^b \ell^c m^d ,
 \quad\quad  \Psi_1 = -C_{abcd} \ell^a n^b \ell^c m^d , \]
\[ \Psi_2  =  -\frac{1}{2} \left( C_{abcd} \ell^a n^b \ell^c n^d
- C_{abcd} \ell^a n^b m^c \bar m^d \right)  , \]
\begin{equation} \Psi_3 = C_{abcd} \ell^a n^b n^c \bar m^d ,  \quad\quad
\Psi_4 = -C_{abcd} n^a \bar m^b n^c \bar m^d, \label{weyl_np}
\end{equation}

\noindent and the Ricci tensor components are

\begin{eqnarray} \Phi_{00} = -\frac{1}{2} R_{ab} \ell^a \ell^b ,
\quad\quad & \Phi_{10} = -\frac{1}{2} R_{ab} \ell^a \bar m^b , &
\quad\quad \Phi_{20} = -\frac{1}{2} R_{ab} \bar m^a \bar m^b ,
\nonumber \\
\Phi_{01} = -\frac{1}{2} R_{ab} \ell^a m^b , \quad\quad &
\Phi_{11} = -\frac{1}{4} \left( R_{ab} \ell^a n^b + R_{ab} m^a
\bar m^b \right) , & \quad\quad \Phi_{21} = -\frac{1}{2} R_{ab}
n^a \bar m^b ,
\nonumber \\
\Phi_{02} = -\frac{1}{2} R_{ab} m^a m^b , \quad\quad & \Phi_{12} =
-\frac{1}{2} R_{ab} n^a m^b , & \quad\quad \Phi_{22} =
-\frac{1}{2} R_{ab} n^a n^b , \nonumber \\ & \Lambda =
\frac{1}{24} R. \label{Ricci_np} \end{eqnarray}


\end{document}